# A Unified Diode Equation for Organic Photovoltaic Devices

Oskar J. Sandberg[1*] and Ardalan Armin[1*]

[1]Sustainable Advanced Materials (Sêr-SAM), Centre for Integrative Semiconductor Materials (CISM), Department of Physics, Swansea University Bay Campus, Swansea SA1 8EN, United Kingdom

*Email: o.j.sandberg@swansea.ac.uk ; ardalan.armin@swansea.ac.uk

**Abstract**

Organic photovoltaics (OPVs) are promising candidates for future sustainable technologies, including applications within the renewable energy sector, such as solar cells and indoor light recycling, and photodetection. However, the performance of OPVs is still inferior compared to established solar cell technologies, partially due to the intrinsically low charge carrier mobilities and large recombination losses of low-permittivity, molecular organic semiconductor systems. To better understand these losses, analytical diode models capable of capturing the underlying device physics are imperative. However, previously proposed analytical models have neglected the effects of injected charge carriers, which is the predominant source for bimolecular recombination in thin-film systems with ohmic contacts. In this work, we derive a unified diode equation for OPVs. Based on a regional approximation approach, we derive an analytical model for the current accounting for the interplay between charge carrier extraction, injection, and bimolecular recombination in organic solar cells. The analytical model is validated by numerical simulations and experimental data. Our findings provide key insights into the mechanisms driving and limiting charge collection, and ultimately the power conversion efficiency, in OPV devices. The presented framework is material-agnostic and generally applicable to sandwich-type thin-film photovoltaic devices, including photodiodes and indoor light-harvesting cells.

## 1. INTRODUCTION

Organic semiconductor-based photovoltaics (OPVs) have shown great promise as an emerging and sustainable solar cell technology [1,2]. In addition, OPVs are suitable for low-light-intensity applications such as indoor light harvesting and photodetection [3,4]. However, OPVs suffer from low charge carrier mobilities and relatively large non-radiative recombination losses. This generally limits the magnitude of the collected photo-induced current and, ultimately, the power conversion efficiency (PCE) of OPV-based solar cells [2,5]. To enhance the device performance of OPVs, all loss mechanisms limiting the current need to be understood and minimized, necessitating a comprehensive understanding of the underlying device physics. In this regard, quantitative analytical diode models which relate the current to key material parameters are invaluable. However, owing to the convoluted and non-linear nature of the physical processes underpinning charge collection in OPVs, an analytical diode model that accurately captures the relevant device physics in these systems has remained elusive.

The typical OPV structure constitutes a thin active layer (often on the order of 100 nm), sandwiched between hole-collecting anode and electron-collecting cathode contact layers. In state-of-the-art OPV cells, the active layer is composed of a blend of electron-accepting (acceptor) and electron-donating (donor) organic semiconductors [2,6]. Under illumination, free electrons and holes are generated in the acceptor and donor phase, respectively, being the result of photo-induced excitons dissociating at donor-acceptor interfaces. The electrons (holes) are then driven toward the cathode (anode) contact by the built-in asymmetry induced by the work function difference of the electrodes. On the way to the respective charge collecting contact, however, a free electron (hole) has a finite probability to encounter a free hole (electron) at a donor-acceptor interface and recombine. This recombination process is generally of a bimolecular nature [7,8], involving the reformation and the subsequent recombination of charge-transfer states or excitons. This latter process is believed to be the leading source of non-radiative recombination in organic solar cells [9,10,11].

Because of the low carrier mobilities, the current-voltage (*J-V*) relationship in OPVs critically depends on the competition between the processes of charge extraction and bimolecular recombination of photogenerated electrons and holes [12,13,14]. This competition is further complicated by the inevitable presence of injected charge carriers diffused into the active layer from the contacts [15,16]. Injected carriers are characterized by highly non-uniform, voltage-dependent carrier profiles and have been found to cause both energy level bending, screening the electric field inside the active layer [17,18], and first-order recombination losses, manifesting as non-ideal Fill Factors (FFs) regardless of light intensity regime [19,20]. While reproduced by numerical simulations, it has been challenging to understand and model these processes analytically due to the multi-dimensional parameter spaces and the non-linear nature of underpinning kinetic processes [21]. Nonetheless, analytical models for the *J-V* behaviour of OPVs have been proposed in the past [22,23,24]. However, these approximations typically assume simplified situations which neglect effects of injected carriers. To our knowledge, an explicit diode equation fully describing the interplay between charge carrier extraction, injection, and bimolecular recombination in low-mobility OPVs has yet to be established.

In this work, we derive a diode equation describing the *J-V* characteristics of thin-film OPV devices. Using a regional approximation approach, expressions for the current are obtained which, for the first time, fully accounts for the interplay between charge carrier extraction, injection, and bimolecular recombination in sandwich-type thin-film devices. The derived approximations are validated by numerical drift-diffusion simulations and applied to experimental results. Further, the obtained findings provide valuable insight into the mechanisms driving and limiting charge collection in OPVs. Additionally, the presented analytical framework provides a figure-of-merit which parametrises *J-V* curves more accurately than previous models, useful for fitting experimental data to extract key material properties. Such parametrisation may also find applications in labelling input *J-V* data for training machines used for device optimisation [25,26]. The developed theoretical framework is material-agnostic and applicable to thin-film photovoltaic devices based on low-mobility semiconductors in general.

## 2. THEORETCIAL BACKGROUND

For a given applied voltage $V$ across the active layer, the total current density $J$ of a thin-film diode device is determined by the flow of electrons and holes within the device. Under steady-state conditions, $J = J_n(x) + J_p(x)$, where $J_n(x)$ and $J_p(x)$ is the electron and hole current density, respectively, at any position $x$ inside the active layer. The electron and hole current densities are related to the photogeneration rate $G$ and recombination rate $R$ of charge carriers within the active layer via the carrier continuity equations:

$$-\frac{1}{q}\frac{\partial J_n}{\partial x} = \frac{1}{q}\frac{\partial J_p}{\partial x} = G(x) - R(x), \tag{1}$$

where $q$ is the elementary charge and $0 < x < d$, with $d$ being the active layer thickness, assuming an active layer sandwiched between an anode ($x = 0$) and a cathode ($x = d$) contact. Further, assuming an effective medium description, $J_n(x)$ and $J_p(x)$ depend on their respective electron and hole densities $n(x)$ and $p(x)$ through the drift-diffusion relations [27,28]:

$$J_n(x) = \mu_n n(x) \frac{\partial E_{Fn}}{\partial x} = \mu_n \left[ q n(x) F(x) + kT \frac{\partial n}{\partial x} \right], \tag{2}$$

$$J_p(x) = \mu_p p(x) \frac{\partial E_{Fp}}{\partial x} = \mu_p \left[ q p(x) F(x) - kT \frac{\partial p}{\partial x} \right]. \tag{3}$$

Here, $\mu_n$ ($\mu_p$) is the electron (hole) mobility, $E_{Fn}$ ($E_{Fp}$) the quasi-Fermi level of electrons (holes), $F$ the electric field, $k$ the Boltzmann constant, and $T$ the absolute temperature. The classical Einstein relation for the electron (hole) diffusion constant, $D_{n(p)} = \mu_{n(p)} kT/q$, has been assumed in the second equality of Eq. (2) (Eq. (3)).

The electric field inside the active layer depends on the densities of free charge carriers through the Poisson equation. In case of an undoped and trap-free active layer, the Poisson equation is given by

$$\frac{\partial F}{\partial x} = \frac{q}{\varepsilon \varepsilon_0} [p(x) - n(x)], \tag{4}$$

with $\varepsilon$ being the relative permittivity of the active layer, and $\varepsilon_0$ the vacuum permittivity. The electric field is related to the electrical potential $\phi(x)$ inside the active layer via $F(x) = -\partial \phi/\partial x$; in this work we define $\phi(x) = -\int_0^x F(x')\,dx'$ with $\phi(0) = 0$ as the reference potential level. Note that $\phi(x)$ is proportional to the effective conduction level $E_c(x)$ in the acceptor and valence level $E_v(x)$ in the donor as $q[\phi(x) - \phi(0)] = E_v(0) - E_v(x)$.

Furthermore, the bulk recombination of charge carriers is assumed to be bimolecular, with the net recombination rate being of the form

$$R(x) = \beta n(x) p(x) - \beta n_i^2, \tag{5}$$

where $\beta$ is the bimolecular recombination rate coefficient. Here, the term $\beta n_i^2$ defines the thermal (dark) generation rate of free charge carriers within the active layer, where $n_i^2 = N_c N_v \exp(-E_g/kT)$ with $N_c$ ($N_v$) being the effective density of conduction (valence) level and $E_g = E_c - E_v$ the effective transport level gap. Since the generation and recombination of free charge carriers in OPVs take place via CT states and excitons [2,29,30,31,32], the corresponding charge carrier rate $G$ and recombination coefficient $\beta$ in Eq. (1) and Eq. (5), respectively, depend on exciton and CT state kinetics [33]. In this picture, $\beta = (1 - P_{CT})\beta_0$ where $\beta_0$ is the rate constant for free electrons and holes to encounter and form bound CT states at D-A interfaces in the bulk [34], and $1 - P_{CT}$ represents the subsequent probability for CT states to decay to the ground state either directly or indirectly (e.g. after back-transfer to excitons) [33]. $\beta$ is also commonly compared to the Langevin recombination constant $\beta_L = q[\mu_n + \mu_p]/(\varepsilon \varepsilon_0)$ [35], considered to be the upper limit for $\beta$ in OPV systems.

Under these conditions, the total current density can be expressed as

$$J = -J_{\text{gen}} + q\beta \int_0^d [n(x)p(x) - n_i^2]\,dx + J_n(0) + J_p(d), \tag{6}$$

where $J_{\text{gen}} \equiv q \int_0^d G(x)\,dx$ is the saturated photogeneration current density. In this work, we consider contacts that are ideally selective for the extraction of majority carriers (electrons at cathode, holes at anode). In other words, majority carriers are assumed to remain at thermal equilibrium at the contacts, such that $p(0) = p_{an}$ and $n(d) = n_{cat}$, while $J_n(0) = J_p(d) = 0$ (no surface recombination of minority carriers). Here, $p_{an}$ ($n_{cat}$) is the corresponding thermal equilibrium density for holes (electrons) at the anode (cathode). Finally, $V$ is related to $F(x)$ via

$$V - V_{bi,0} = \int_0^d F(x)\,dx, \tag{7}$$

where $qV_{bi,0} = kT \ln(p_{an} n_{cat}/n_i^2)$ corresponds to the work function difference between the anode and cathode contact.

To evaluate $J$ for a given $V$, the set of coupled differential equations Eqs. (1)-(5) needs to be solved for $\phi(x)$, $n(x)$ and $p(x)$. In the next section, we will present an analytical framework for $J$ based on approximative solutions of $\phi(x)$, $n(x)$ and $p(x)$. To validate the obtained approximate expressions, we will compare them with the results of numerical simulations representing the "exact" solutions. For this purpose, we use a previously established numerical device model (see Ref. [36]), based on the Scharfetter-Gummel discretization scheme and Gummel's iteration method [37,38,39]. In the simulations, we consider an active layer having $d = 100$ nm, $N_v = N_c = 10^{21}$ cm$^{-3}$, and an energy level gap of $E_g = 1.42$ eV. Further, we assume a uniform photogeneration rate ($G(x) = G$) that scales linearly with light intensity and takes a value of $6.24 \times 10^{21}$ cm$^{-3}$s$^{-1}$ under 1 sun conditions. Unless otherwise stated, balanced mobilities ($\mu = \mu_n = \mu_p$) will be assumed in the simulations. Finally, we will assume that the contacts are ohmic, corresponding to injecting contacts at which the carrier density is high enough (negligible resistance) not to limit the majority-carrier current. To ensure this condition, we assume $p_{an} = N_v$ and $n_{cat} = N_c$.

## 3. RESULTS AND DISCUSSION

In Fig. 1 the effect of the interplay between charge collection and bimolecular recombination on the *J-V* characteristics is demonstrated. A photovoltaic device with $\mu = 10^{-4}$ cm$^2$/Vs is simulated and compared to the idealized case with perfect charge collection ($\mu \to \infty$), assuming a fixed $\beta = 10^{-10}$ cm$^3$/s. Fig. 1(a) and (b) shows the normalized currents $J/J_{\text{gen}}$ under 1 sun (solar cell) and 0.001 suns (relevant for indoor PV) illumination conditions, respectively. Independent of the light intensity, decreasing $\mu$ generally reduces both the short-circuit current density ($J_{SC}$) and the FF. This suggests that, even at low light intensities, there can be a substantial charge collection loss caused by bimolecular recombination in case of low mobilities. Conversely, in the limit of high mobilities, both $J_{SC}$ and FF saturate to their maximum values (set by the prevailing intensity) as $J_{SC} \to J_{gen}$. Note that the open-circuit voltage ($V_{OC}$) is independent of $\mu$ in this case (since $\beta$ is fixed) and only depends on the light intensity. For an analytical model of the *J-V* characteristics to be accurate, it is imperative that it reproduces the mobility and intensity dependent behaviours in Fig. 1. In the following, approximations relating the current and bimolecular recombination in thin-film OPV devices with ohmic contacts will be derived.

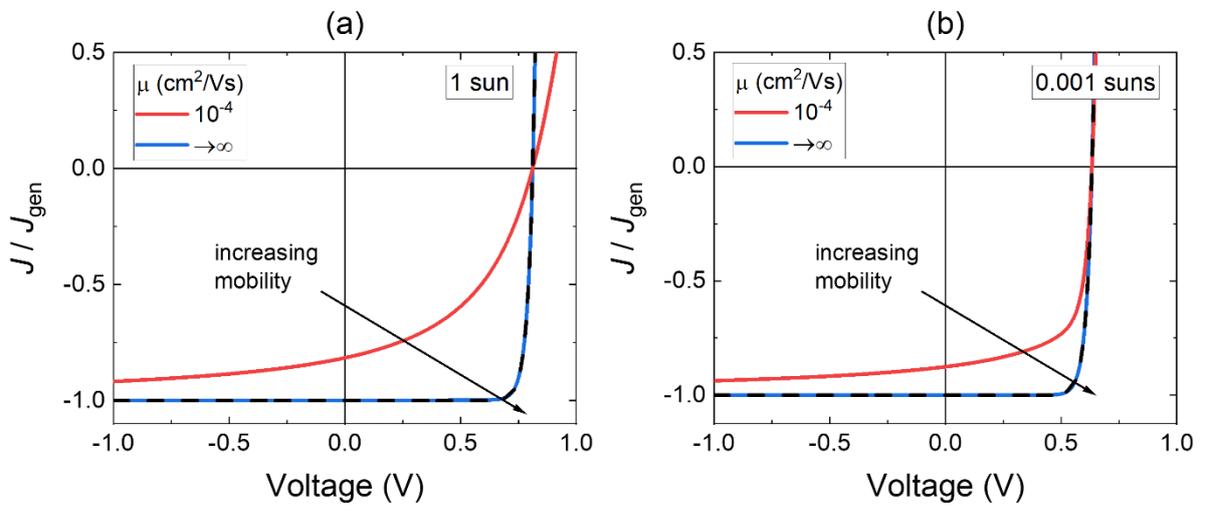

**Figure 1.** Simulated *J-V* characteristics of an OPV device under (a) 1 sun and (b) 0.001 suns illumination conditions demonstrating the effect of finite mobilities on the current density. The case of balanced mobilities with $\mu = 10^{-4}$ cm²/Vs (red line) is shown and compared to the idealized case of $\mu \to \infty$ (blue line). A fixed bimolecular recombination coefficient $\beta = 10^{-10}$ cm³/s is assumed for both cases. The predictions based on the ideal diode equation Eq. (17) are indicated by black dashed lines.

### A. Analytical approach

To obtain approximate solutions for the charge carrier transport equations, we use the following ansatz for the electron and hole density:

$$n(x) = n_0(x) + \Delta n(x), \tag{8}$$

$$p(x) = p_0(x) + \Delta p(x). \tag{9}$$

Here, $n_0(x)$ and $p_0(x)$ are defined as the respective ideal carrier densities expected in case of flat quasi-Fermi levels ($\partial E_{Fn}/\partial x = \partial E_{Fp}/\partial x = 0$) separated by $qV$. In accordance with Eq. (2) and (3), we have by definition that

$$qn_0(x)F(x) = -kT\frac{\partial n_0(x)}{\partial x}, \tag{10}$$

$$qp_0(x)F(x) = kT\frac{\partial p_0(x)}{\partial x}, \tag{11}$$

for every $x$ inside the active layer. Noting that $J = J_n(x) = J_p(x) = 0$ at open circuit, it directly follows that $n(x) = n_0(x)$ and $p(x) = p_0(x)$ at $V = V_{OC}$. In other words, $n_0(x)$ and $p_0(x)$ at any given applied voltage, $V$ in the dark, are identical to the respective $n(x)$ and $p(x)$ obtained at open circuit under a light intensity that results in a $V_{OC}$ equal to $V$.

Conversely, $\Delta n(x)$ and $\Delta p(x)$ represent the corresponding deviations of $n(x)$ and $p(x)$ from their ideal carrier densities. $\Delta n(x)$ and $\Delta p(x)$ are determined by the charge carrier transport properties inside the device. Making use of Eq. (10) and (11), in conjunction with Eq. (2) and (3), in Eq. (1) the electron and hole continuity equations can be expressed as

$$\mu_n F\frac{\partial \Delta n}{\partial x} + \mu_n \Delta n\frac{\partial F}{\partial x} + \mu_n \frac{kT}{q}\frac{\partial^2 \Delta n}{\partial x^2} = -G + \beta\left[p_0 n_0 - n_i^2 + p_0\Delta n + n_0\Delta p + \Delta n\Delta p\right], \tag{12}$$

$$\mu_p F\frac{\partial \Delta p}{\partial x} + \mu_p \Delta p\frac{\partial F}{\partial x} - \mu_p \frac{kT}{q}\frac{\partial^2 \Delta p}{\partial x^2} = G - \beta\left[p_0 n_0 - n_i^2 + p_0\Delta n + n_0\Delta p + \Delta n\Delta p\right]. \tag{13}$$

While general analytically tractable solutions to Eq. (12) and (13) cannot be found, approximate expressions can be obtained depending on the dominant recombination terms. Note that $\Delta n$ and $\Delta p$ should primarily be considered mathematical tools which may take positive or negative values. In regions where $\Delta n \gg n_0$ ($\Delta p \gg p_0$) apply, however, $\Delta n$ ($\Delta p$) may be identified as the excess density of photogenerated electrons (holes).

*1. Ideal charge collection*

We first consider the idealized zeroth order case, when $n(x) \to n_0(x)$ and $p(x) \to p_0(x)$. Excluding open-circuit conditions, this corresponds to the limit of infinite mobilities (or ideal conductivities). This follows from the fact that for $J_n(x)$ and $J_p(x)$ to remain finite as $\mu_n$ and $\mu_p$ approaches infinity, we must generally have $\frac{\partial E_{Fn}}{\partial x} = \frac{\partial E_{Fp}}{\partial x} \to 0$. Fig. 2(a) and (b) show the carrier densities inside the active layer (under 1 sun illumination condition) for different mobilities at $V = 0$ and $V = 0.5$ V, respectively. The corresponding energy level diagrams are depicted in Fig. 2(c) and (d). As expected, $E_{Fn}$ and $E_{Fp}$ in the limit of high mobilities are flat throughout the active layer and separated by $qV$ In other words, the free carriers are in electrochemical equilibrium with their collecting electrodes.

As $n(x) \to n_0(x)$ and $p(x) \to p_0(x)$, the recombination rate is dominated by the $\beta p_0(x)n_0(x)$ term in Eq. (12) and (13) (i.e., the zeroth order term with respect to photogenerated carriers). This term can be directly evaluated noting that $n_0(x)$ and $p_0(x)$, as per Eq. (10) and (11), can be equivalently expressed as

$$n_0(x) = n_{cat} \exp\left(\frac{q[V-V_{bi,0}+\phi(x)]}{kT}\right), \tag{14}$$

$$p_0(x) = p_{an} \exp\left(-\frac{q\phi(x)}{kT}\right). \tag{15}$$

As a result, in this limit $n_0(x)p_0(x)$, and thus the bimolecular recombination rate, is independent of the position inside the active layer. Making use of the definition of $V_{bi,0}$, the generalised mass action law is further obtained as:

$$n_0(x)p_0(x) = n_i^2 \exp\left(\frac{qV}{kT}\right). \tag{16}$$

Hence, for selective contacts ($J_n(0) = J_p(d) = 0$), the total current density [Eq. (6)] in the limit of ideal charge collection ($\mu_{n,p} \to \infty$) is given by

$$J_{\text{ideal}}(V) = -J_{\text{gen}} + J_0 \left[\exp\left(\frac{qV}{kT}\right) - 1\right], \tag{17}$$

where $J_0 = q\beta n_i^2 d$ is the dark saturation current density induced by thermal generation of charge carriers in the active layer. As expected, in the high-mobility limit, $J(V)$ is of an identical form to the ideal diode equation. Indeed, Eq. (17) (indicated by the dashed lines) precisely reproduces the simulated J-V curves when $\mu \to \infty$ in Fig. 1.

### 2. Regional approximation

Based on Eq. (14) and (15), expressions for the ideal carrier densities can be derived by making use of a regional approximation. In the ideal limit, the density of electrons and holes dominate at cathode and anode contact, respectively, while decreasing exponentially away from the contact (see Fig. 2a,b). Subsequently, the active layer can be divided into a hole-dominated ($0 < x < x^*$) and an electron-dominated region ($x^* < x < d$), where $x^*$ is the width of the hole-dominated region. Assuming that $p_0 \gg n_0$ in the hole-dominated region, analytical approximations for $p_0(x)$ can be obtained [40,41]. For $x < x^*$, Eq. (4) then simplifies as

$$\frac{\partial F(x)}{\partial x} = -\frac{\partial^2 \phi(x)}{\partial x^2} \approx \frac{q p_0(x)}{\varepsilon \varepsilon_0}. \qquad (18)$$

Substituting Eq. (15) into Eq. (18) and solving for $\phi(x)$ yields

$$\phi(x) = \frac{2kT}{q} \ln\left(\frac{2kT}{qF_0 \lambda_{an}} \sinh\left[\frac{qF_0 x}{2kT} + \sinh^{-1}\left(\frac{qF_0 \lambda_{an}}{2kT}\right)\right]\right), \qquad (19)$$

for $F_0 \gg 2kT/qd$, where $\lambda_{an} = \sqrt{2\varepsilon\varepsilon_0 kT/[q^2 p_{an}]}$ represents the Debye screening length for holes at the anode contact and $F_0$ is an integration constant of the electric field. Thus, the electric field in this region is obtained as

$$F(x) = -F_0 \coth\left[\frac{qF_0 x}{2kT} + \sinh^{-1}\left(\frac{qF_0 \lambda_{an}}{2kT}\right)\right]. \qquad (20)$$

Note that sufficiently far away from the contact, $F \to -F_0$ in Eq. (20). Concomitantly, $F_0$ represents the magnitude of the electric field deep inside the active layer. A completely analogous treatment applies for $x > x^*$ (assuming $n_0 \gg p_0$) [42]. Demanding that $F(x)$ is continuous at $x = x^*$, it can be shown that $x^* \approx d/2$ for ohmic contacts, corresponding to $\lambda_{an}, \lambda_{cat} \ll d$, where $\lambda_{cat}$ is the corresponding Debye screening length for electrons at the cathode contact. Finally, an expression for $F_0$ can be obtained after matching $\phi(x)$ at $x = x^*$ and applying the boundary condition Eq. (7).

In general, $F_0$ depends on the contact properties (represented by $\lambda_{an}$, $\lambda_{cat}$ and $V_{bi,0}$) and the applied voltage through a transcendental dependence. In case of ohmic contacts ($\lambda_{an}, \lambda_{cat} \ll d$), however, $F_0$ may be approximated as [42]

$$F_0 \approx \frac{1}{d}\left[V_{bi} - V \times \left(1 + \frac{\ln|2|}{\frac{qV_{bi}}{8kT} - 1}\right)\right], \qquad (21)$$

for $V_{bi} - V > 8\ln(2)\, kT/q$, where $V_{bi}$ is the effective built-in voltage inside the active layer at $V = 0$, determined via

$$V_{bi} = V_{bi,0} - \frac{2kT}{q} \ln\left(\frac{q^2 d^2 \sqrt{p_{an} n_{cat}}}{2\varepsilon\varepsilon_0 kT}\right) + \frac{4kT}{q} \ln\left(\frac{qV_{bi}}{kT}\right). \qquad (22)$$

The final term within the round brackets on the right-hand side of Eq. (21) may be viewed as a correction term accounting for the fact that the effective built-in voltage inside the active layer depends weakly on the applied voltage.

In Fig. 2, $p_0(x)$ predicted by Eq. (15) (dashed lines), with $\phi(x)$ given by Eq. (19), are compared with the simulated $p(x)$ (solid lines). Indeed, Eq. (15) reproduces the simulated $p(x)$ in the limit $\mu_p \to \infty$. Inside the active layer, Eq. (19) predicts $\phi(x)$ to be linearly dependent on $x$, translating into exponentially varying carrier densities in this limit. Near the contacts, however, the accumulation of majority carriers induces significant energy level bending resulting in non-exponential behaviour within these regions. It should be noted that, although initially derived for $x < x^*$, the approximation based on Eq. (19) generally applies beyond this region; in fact, except for the electron-accumulation region near the cathode contact, Eq. (19) remains a good approximation for $\phi(x)$ throughout the entire active layer.

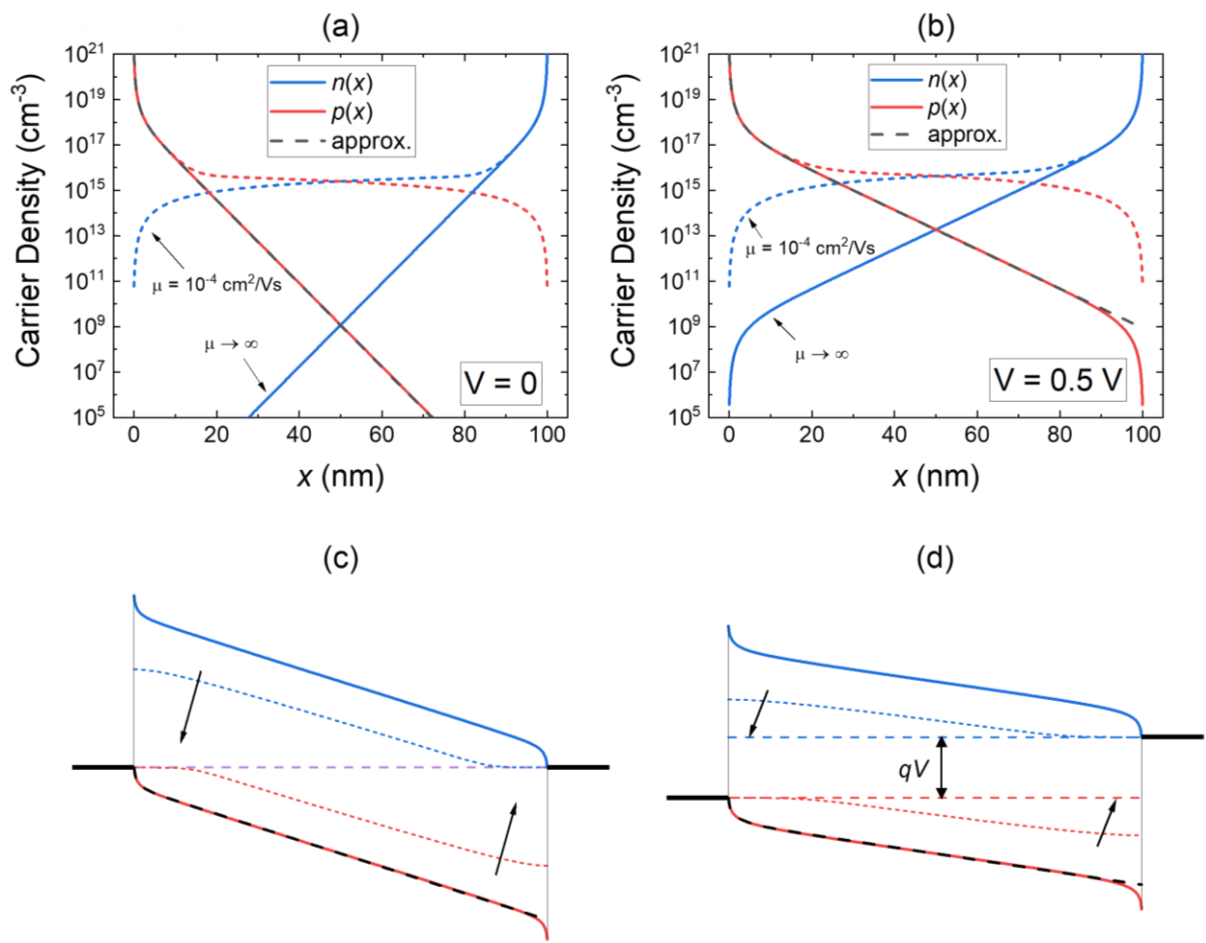

**Figure 2.** Simulated carrier density profiles of the OPV device from Fig. 1 at 1 sun for (a) $V = 0$ and (b) $V = 0.5$ V. The solid and dotted lines indicate the case of $\mu \to \infty$ and $\mu = 10^{-4}$ cm²/Vs, respectively. In (c) and (d), the corresponding energy level diagrams for $V = 0$ and $V = 0.5$ V are shown, respectively. The conduction and valence levels are indicated by blue and red solid lines, respectively, while the electron and hole quasi-Fermi levels are indicated by blue and red dashed lines. For comparison, the analytical approximations based on Eq. (15) and Eq. (19) are depicted by the black dashed lines.

In case of finite mobilities, Eq. (15) is also seen to accurately describe $p(x)$ within the anode contact region; this is a consequence of the high hole density resulting in a virtually flat $E_{Fp}$ [cf. Eq. (3)], and thus $p(x) = p_0(x)$, within this region. Subsequently, the majority carrier densities within the contact regions are independent of charge transport properties, consistent with the defining characteristic of ohmic contacts. Finally, the $E_v(x)$ predicted based on Eq. (19) (dashed line) are compared with the simulated $E_v(x)$ (solid lines) in Fig. 2(c) and (d); an excellent agreement is obtained. It should be noted that the simulated $E_v(x)$ at the two different mobility cases in Fig. 2(c) and (d) coalesce, suggesting that $\phi(x)$ is independent of mobilities in this case.

**B. First-order approximation**

In case of finite mobilities $n(x) \approx n_0(x)$ and $p(x) \approx p_0(x)$ only within the high-density region near the cathode and anode contact, respectively, while the carrier densities strongly deviate from the ideal limit well inside the active layer, as shown in Fig. 2(a) and (b). The deviations $\Delta n(x)$ and $\Delta p(x)$ depend on both the light intensity and the charge transport properties. In the low-intensity limit, however, when the second-order recombination term $\beta \Delta n \Delta p$ is negligible (the recombination rate is only determined by the 1st order terms) and the space charge induced by $\Delta n$ and $\Delta p$ is insignificant (i.e., $F(x)$ is given by Eq. (20)), approximative solutions to Eq. (12) and (13) can be obtained.

*1. Drift-only solutions for $\Delta n(x)$*

We consider the drift-only case, assuming the diffusion component of $\Delta n(x)$ to be negligible. For electrons sufficiently far away from the cathode contact, the term $n_0(x)\Delta p(x)$ can be neglected. Then, noting that $p_0(x)$ is related to $F(x)$ via Eq. (18), in this region, Eq. (12) reduces to

$$-\mu_n F(x)\frac{\partial \Delta n}{\partial x} + [z_n - 1]\mu_n \frac{\partial F}{\partial x}\Delta n(x) = G_R, \qquad (23)$$

where $G_R = G - \beta n_i^2[\exp(qV/kT) - 1]$ and $z_n = \beta/\beta_n$, with $\beta_n \equiv q\mu_n/(\varepsilon\varepsilon_0)$ being an equivalent electron-only Langevin rate constant. The solution for Eq. (23) is of the form:

$$\Delta n(x) = \frac{G_R}{\mu_n F_0} \times \frac{\int_0^x \tanh^{z_n}\left[\frac{qF_0 x'}{2kT}+\sinh^{-1}\left(\frac{qF_0 \lambda_{an}}{2kT}\right)\right]dx'}{\tanh^{z_n-1}\left[\frac{qF_0 x}{2kT}+\sinh^{-1}\left(\frac{qF_0 \lambda_{an}}{2kT}\right)\right]}, \qquad (24)$$

with $\Delta n(0) = 0$ and using Eq. (20) for $F(x)$. An analogous expression can be obtained for $\Delta p(x)$ sufficiently far away from the anode contact.

For the case of perfect ohmic contacts (corresponding to $p_{an} \to \infty$ and $\lambda_{an} \to 0$), $\Delta n(x)$ far away from both contacts simplifies as

$$J_n(x) \approx -q\mu_n F_0 \Delta n(x) \approx -qG_R[x - x_{0n}], \qquad (25)$$

where $x_{0n} = (kT/qF_0) \times f_0(\beta/\beta_n)$ with $f_0(z) \equiv 2 \times \int_0^\infty 1 - \tanh^z(u)\, du$. The integral function $f_0(z)$ can be readily evaluated for special cases (e.g., $f_0(2) = 2$, $f_0(1) = \ln(4)$); for a general $z$, we find

$$f_0(z) = \psi\left(\frac{z+1}{2}\right) + \gamma + \ln(4), \qquad (26)$$

where $\psi$ is the digamma function and $\gamma \approx 0.5772$ is the Euler constant. For $z \ll 1$ we expect $f_0(z) \to \pi^2 z/4$, while $f_0(z) \approx \ln[2z \exp(\gamma)]$ in the limit of large $z$. For practical purposes, however, it is convenient to assume the following numerical approximation for Eq. (26):

$$f_0(z) \approx \ln\left(\sqrt{1 + B^2 z^2} + \frac{\pi^2 z}{4}\right), \qquad (27)$$

with $B = [8\exp(\gamma) - \pi^2]/4 \approx 1.0947$, which reproduces the two extreme cases.

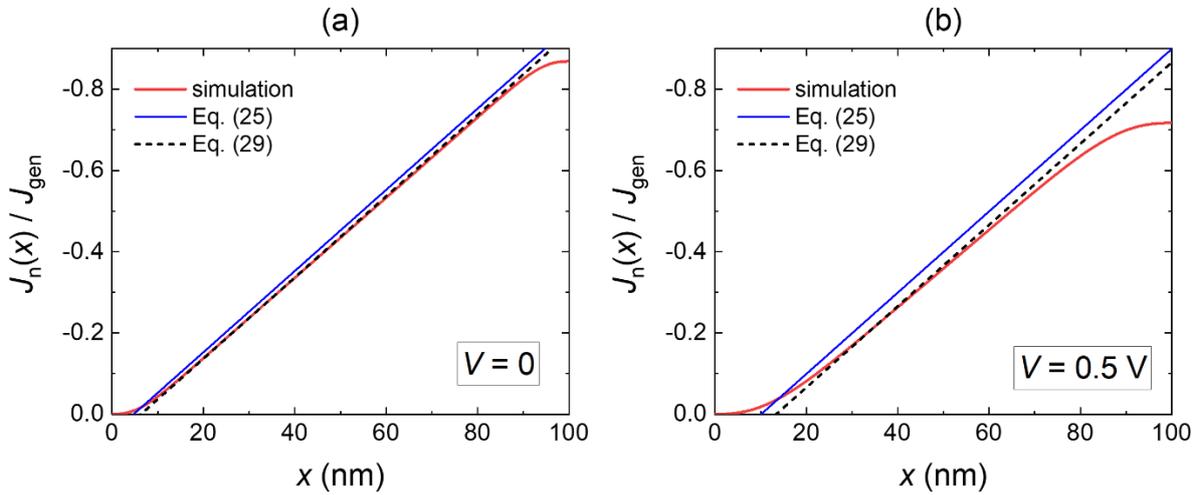

**Figure 3.** Normalized electron current density profiles simulated at (a) short circuit and (b) $V = 0.5$ V for the low-mobility device from Fig. 1(b) ($\mu = 10^{-3}$ cm²/Vs and $\beta = \beta_L$), indicated by the red solid lines. The corresponding approximations Eq. (25) and Eq. (29) for the electron current densities well inside the active layer are indicated by the solid blue lines and dashed lines, respectively.

## 2. Accounting for diffusion

The above analysis culminating in Eq. (24) predicts that $J_n(x)$ is given by Eq. (25) far away from the anode contact ($x \gg x_{0n}$). On the other hand, for $x < x_{0n}$, Eq. (24) approximates as

$$\Delta n(x) \approx \frac{qG_R x^2}{2\mu_n kT[1+\beta/\beta_n]}, \tag{28}$$

for perfect ohmic contacts ($\lambda_{an} \to 0$). However, this treatment neglects diffusion of $\Delta n(x)$, assuming the drift component $\Delta J_{n,\text{drift}} = q\mu_n \Delta n F$ to be much larger than the corresponding diffusion component, $\Delta J_{n,\text{diff}} = \mu_n kT(\partial \Delta n/\partial x)$. While this is a good approximation for $x \gg x_{0n}$, the assumption is generally not valid for electrons close to the anode contact. Using Eq. (28), the corresponding $\Delta J_{n,\text{diff}}$ is approximately equal but opposite to the drift component ($\Delta J_{n,\text{diff}} \approx -\Delta J_{n,\text{drift}}$), suggesting the actual total electron current ($J_n = \Delta J_{n,\text{drift}} + \Delta J_{n,\text{diff}}$) for $x < x_{0n}$ to be close to zero. Therefore, to a first approximation, the drift-only analysis underestimates the magnitude of $J_n(x)$ by $\Delta J_{n,\text{drift}}(x_{0n})$ for $x \gg x_{0n}$. Based on Eq. (28), we expect this correction to be given by $\Delta J_{n,\text{drift}}(x_{0n}) \approx -qG_R x_{0n}/(z_n + 1)$. A completely analogous consideration applies for $\Delta p(x)$.

Based on these considerations, the electron and hole current densities well inside the active layer can then be approximated as

$$J_n(x) = -qG_R[x - x_{Rn}], \tag{29}$$

$$J_p(x) = qG_R[x - d + x_{Rp}]. \tag{30}$$

where $x_{Rn} = \frac{kT}{qF_0} \times f_R(\beta/\beta_n)$ and $x_{Rp} = \frac{kT}{qF_0} \times f_R(\beta/\beta_p)$, with $\beta_p = q\mu_p/(\varepsilon\varepsilon_0)$ being an equivalent hole-only Langevin rate constant. Further, after approximating $f_0(z)$ by Eq. (27), $f_R(z)$ is given by

$$f_R(z) = \frac{z+2}{z+1} f_0(z) \approx \frac{z+2}{z+1} \ln\left(\sqrt{1 + B^2 z^2} + \frac{\pi^2 z}{4}\right). \tag{31}$$

Fig. 3 shows the simulated $J_n(x)$ of the low-mobility device in Fig. 1(b) at low illumination conditions (where the first-order recombination channel dominates). Comparing the simulations with Eq. (29) a yields a good agreement between the two inside the active layer.

### 3. Approximate expression for the current density

As is implied in Fig. 3, the effect of injected carriers is to induce recombination zones for minority carriers near the contacts. In accordance with Eq. (29) and (30), $x_{Rn}$ and $x_{Rp}$ can be viewed as the corresponding effective widths within which all minority carriers recombine near the anode and cathode contact, respectively. Further, these recombination zones are separated by a recombination-free charge collection zone of width

$$d_{\text{eff}} = d - x_{Rn} - x_{Rp} \tag{32}$$

inside the active layer, as illustrated in the schematic energy level diagram in Fig. 4(a). Note that the widths of the recombination zones depend on the ratios between the mobilities and recombination coefficient and grow with increasing applied voltage. As a result, the recombination-free charge collection zone is reduced with increasing voltage giving rise to an increased charge collection loss. Conversely, in the limit of large reverse-bias voltages, high mobilities, or small $\beta$, we expect $d_{\text{eff}} \to d$, as the ideal high-mobility limit is approached.

Finally, combining Eq. (29) and (30), the total current density $J = J_n(x) + J_p(x)$, evaluated inside the active layer, can be approximated as

$$J(V) = \left(1 - \frac{\theta kT}{qF_0(V)d}\right) \times J_{\text{ideal}}(V), \tag{33}$$

where $J_{\text{ideal}}(V)$ is given by Eq. (17) and

$$\theta = f_R(\beta/\beta_n) + f_R(\beta/\beta_p). \tag{34}$$

with $f_R$ defined through Eq. (31). Subsequently, the effect of a finite mobility is to reduce the magnitude of the current density, relative to the idealized high-mobility limit [Eq. (17)], by a prefactor given by $d_{\text{eff}}/d$. In addition to the applied voltage and the (effective) built-in voltage (via $F_0$), this current reduction factor only depends on $\beta/\beta_n$ and $\beta/\beta_p$ in the active layer (through $\theta$).

Eq. (33) generally describes the current density under low intensity conditions when first-order recombination with injected carriers dominates. In Fig. 4 the light intensity dependence of $J_{\text{SC}}/J_{\text{gen}}$ of the low-mobility case from Fig. 1 ($\mu = 10^{-4}$ cm²/Vs), along with the corresponding *J-V* curves at 0.001 and 1 sun conditions, is simulated and compared to the analytical prediction Eq. (33). Indeed, a nearly perfect agreement between Eq. (33) and the simulations is obtained at low intensities (0.001 suns). Note that for the case $\mu \to \infty$, Eq. (33) reduces back to the ideal diode equation $J_{\text{ideal}}(V)$.

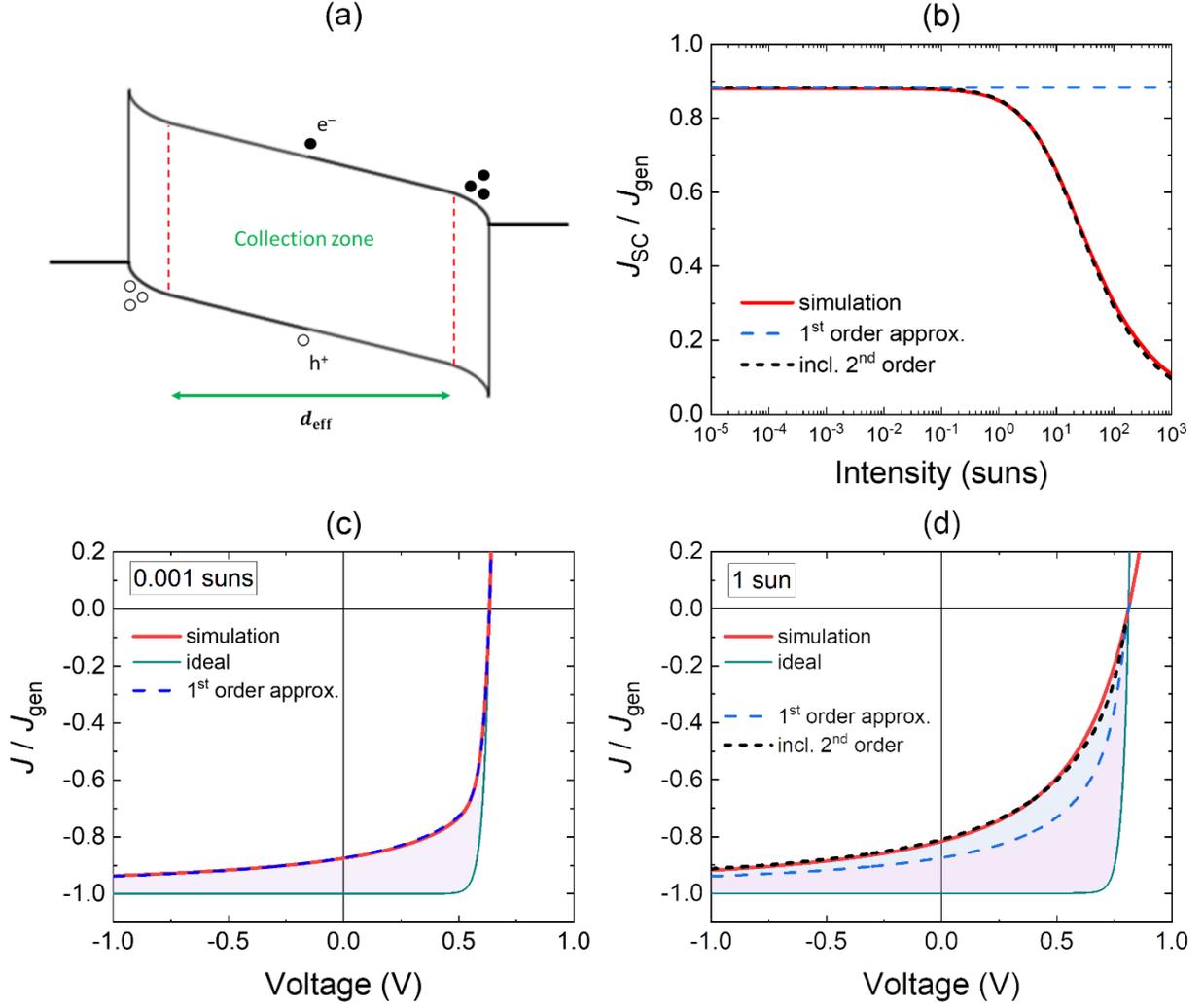

**Figure 4.** In (a) a schematic energy level diagram indicating the recombination zones near the contacts, where first-order recombination between injected and photogenerated carriers dominate, and charge collection zone of width $d_{\text{eff}}$ where first-order recombination is negligible. In (b), the simulated intensity dependence of $J_{\text{SC}}/J_{\text{gen}}$ (red solid line) of the low-mobility device ($\mu = 10^{-4}$ cm$^2$/Vs) from Fig. 1 is shown alongside the corresponding analytical approximations based on Eq. (33) (blue dashed line) and Eq. (41) (black dotted line). In (c) and (d), the corresponding simulated J-V curves at 0.001 suns [from Fig. 1(b)] and 1 sun [from Fig. 1(a)] are shown, respectively, and indicated by red solid lines. The analytical approximations based on Eq. (33) (accounting for 1$^{\text{st}}$ order recombination with injected carriers) and Eq. (41) (also including effects of 2$^{\text{nd}}$ order recombination among photogenerated carriers) are indicated by blue dashed and black dotted lines, respectively. The associated contributions from 1$^{\text{st}}$ and 2$^{\text{nd}}$ order recombination to the total charge collection losses, relative to the ideal limit [Eq. (17), dark-cyan solid lines], are highlighted by purple- and blue-shaded areas, respectively.

## C. Accounting for second-order recombination between photogenerated carriers

As evident from Fig. 4, a deviation between the simulations and the prediction based on Eq. (33) eventually appears at high enough intensities as second-order recombination between photogenerated electrons and holes become important. Under these conditions, the $\beta \Delta n \Delta p$ term in Eq. (12) can no longer be neglected. Subsequently, there will be a competition between charge extraction and second-order recombination taking place within the bulk of the active layer, with $J(V)$ subsequently deviating from Eq. (33). However, a correction to Eq. (33) can be obtained by accounting for second-order recombination within the charge collection zone, $x_{Rn} < x < d - x_{Rp}$. Since $\Delta n \gg n_0$ and $\Delta p \gg p_0$, within this region, $\Delta n$ and $\Delta p$ represent the actual excess densities of electrons and holes induced by photogeneration. First-order recombination (dominating near the contacts) is accounted for by assuming that $J_n(x_{Rn}) = 0$ and $J_p(d - x_{Rp}) = 0$. In other words, the device is effectively treated as having a thickness $d_{\text{eff}}$, where all electrons (holes) photogenerated within $0 < x < x_{Rn}$ ($d - x_{Rp} < x < d$) are lost due to recombination.

### 1. Effect of second-order recombination inside the collection zone

We consider the effect of second-order recombination at small voltages when $V_{\text{OC}} - V \gg kT/q$, corresponding to $G_R \approx G$. Then, assuming that the transport of photogenerated carriers is dominated by drift and that space charge effects are negligible, within the charge collection zone, Eq. (12) simplifies as

$$-\mu_n F \frac{\partial \Delta n}{\partial x} = G - \beta \Delta n(x) \Delta p(x) = G - \frac{\beta J}{q \mu_p F} \Delta n(x) + \frac{\mu_n \beta}{\mu_p} \Delta n(x)^2, \quad (35)$$

where $J = q[\mu_n \Delta n(x) + \mu_p \Delta p(x)]F$ was used in the last step to eliminate $\Delta p(x)$. Assuming that $\Delta n(x_{Rn}) = 0$, while noting that the total current $J$ is independent of $x$, Eq. (35) can be readily solved for $\Delta n(x)$ with the result:

$$\Delta n(x) = \sqrt{\frac{\mu_p G}{\mu_n \beta}} \sqrt{1 - \left(\frac{J}{J_\beta}\right)^2} \tan\left(2\sqrt{1 - \left(\frac{J}{J_\beta}\right)^2} \frac{qG[x - x_{Rn}]}{J_\beta} + \sin^{-1}\left(\frac{J}{J_\beta}\right)\right) + \frac{J}{2q\mu_n F}, \quad (36)$$

for $x_{Rn} < x < d - x_{Rp}$, where $J_\beta = 2q\sqrt{\mu_n \mu_p}\sqrt{(G/\beta)}|F|$. An analogous treatment applies for photogenerated holes within the collection zone; in fact, assuming that $\Delta p = 0$ at $x = d - x_{Rp}$, it can be shown that $\Delta p(x) = (\mu_n/\mu_p) \times \Delta n(d - x_{Rp} - x)$.

Unfortunately, an analytically tractable explicit solution of Eq. (36) in terms of the total current density $J$ cannot be established [24]. However, noting that $\mu_n \Delta n = \mu_p \Delta p$ (and thus $J_n = J_p = J/2$) at the midplane of the charge collection zone, the following implicit relation is obtained:

$$\sin^{-1}\left(\frac{J}{J_\beta}\right) = -\frac{J^*_{\text{gen}}}{J_\beta}\sqrt{1-\left(\frac{J}{J_\beta}\right)^2}. \tag{37}$$

Here, $J^*_{\text{gen}}$ is defined as $J^*_{\text{gen}} = qGd_{\text{eff}}$ and corresponds to the magnitude of the current density obtained when all carriers photogenerated within the collection zone are extracted.

We note that the derivation of Eq. (36) assumes the electric field to be uniform in the charge collection zone. At small voltages, the electric field within this region is well approximated by $F \approx -F_0$. However, the derivation of Eq. (36) also neglects the effect of diffusion of photogenerated carriers induced by super-linear density profiles in the collection zone, resulting in Eq. (37) overestimating the magnitude of the actual $J$. The deviation may be estimated from the difference between the expected diffusion components ($\mu_n kT\, \partial \Delta n(x)/\partial x$) of Eq. (36) evaluated at the edge ($x = x_{Rn}$) and midplane of the collection zone. It turns out that this effect can be corrected for in Eq. (37) by modifying $J_\beta$ as [43]

$$J_\beta \approx 2q\sqrt{\mu_n \mu_p}\sqrt{\frac{G}{\beta}}F_0 \times \left[1 + \frac{2kT}{qF_0 d_{\text{eff}}}\right]^{-1/2}. \tag{38}$$

where the final term on the right-hand side of Eq. (38) represents a second-order correction induced by a diffusion of $\Delta n(x)$ in the collection zone.

In accordance with Eq. (37) the current density depends on the ratio $J^*_{\text{gen}}/J_\beta$, as shown in Fig. 5. As expected, $J$ approaches $-J^*_{\text{gen}}$ in limit when $J^*_{\text{gen}} \ll J_\beta$, corresponding to the low-intensity limit when second-order recombination within the charge collection zone is negligible. At high intensities when $J^*_{\text{gen}} \gg J_\beta$, in turn, Eq. (37) predicts that $J \to -J_\beta$; in this regime, the current is strongly limited by second-order bimolecular recombination between photogenerated electrons and holes inside the collection zone. For the general case, an explicit approximation for $J$ can be obtained based on expanding the left-hand-side of Eq. (37) as $\sin^{-1}(J/J_\beta) \approx J/J_\beta$. Subsequently, the following approximative expression was found:

$$J \approx -\frac{J^*_{\text{gen}}}{\sqrt{1+K(V)}}, \tag{39}$$

with

$$K(V) = \left(\frac{J^*_{\text{gen}}}{J_\beta}\right)^2 + \kappa \frac{J^*_{\text{gen}}}{J_\beta}, \tag{40}$$

where $\kappa$ is a correction factor. Here, the case $\kappa = 0$ corresponds to the first-order approximation $\sin^{-1}(J/J_\beta) \approx J/J_\beta$ in Eq. (37). However, by introducing an additional correction term (with non-zero $\kappa$), the accuracy between the exact value for $J$ based on Eq. (37) and Eq. (39) can be further improved. As shown in Fig. 5, a good agreement is obtained for $\kappa = 0.173$.

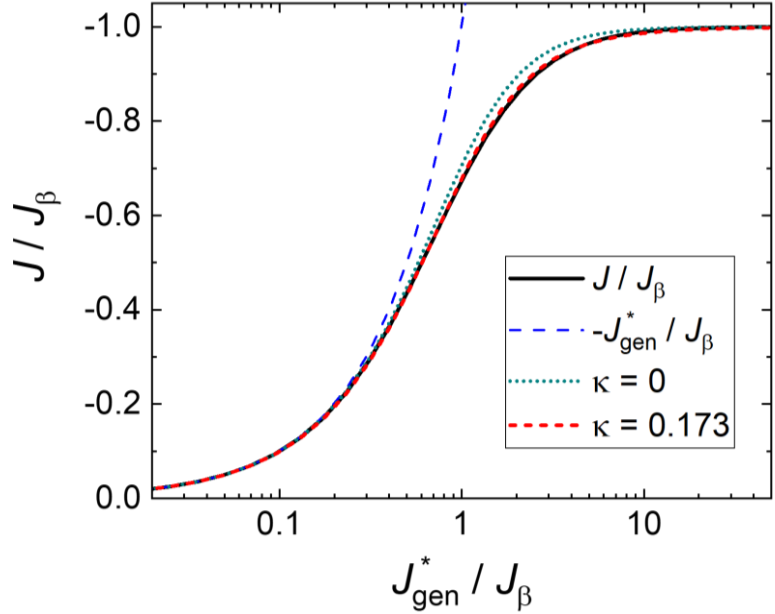

**Figure 5.** Calculated $J/J_\beta$ as a function of $J_{\text{gen}}^*/J_\beta$. The solid line corresponds to the exact (implicit) solution to Eq. (37). The approximations based on Eq. (39) with $\kappa = 0$ and $\kappa = 0.173$ are indicated by dotted and short-dashed lines, respectively. The case $J = J_{\text{gen}}^*$ (long-dashed lines) neglecting second-order recombination between photogenerated carriers has been included for comparison.

### 2. Revised approximation for the current density

Based on Eq. (39), Eq. (33) can be extended to account for the effect of second-order recombination among photogenerated charge carriers. Noting that $J_{\text{gen}}^*$ is equal to Eq. (33) for $V_{\text{OC}} - V \gg kT/q$, it becomes evident that the effect of second-order recombination is to reduce the current density by a factor $1/\sqrt{1 + K(V)}$. Hence, the current density can be approximated as

$$J(V) = \frac{1}{\sqrt{1+K(V)}} \times \left(1 - \frac{\theta kT}{qF_0(V)d}\right) \times J_{\text{ideal}}(V). \tag{41}$$

In other words, apart from the reduction induced by recombination with injected carriers at the contacts, the magnitude of the current density is further reduced, relative to $J_{\text{ideal}}(V)$, by an additional reduction factor induced by second-order recombination among photogenerated carriers within the charge collection zone.

In accordance with Eq. (41), the current loss induced by second-order recombination among photogenerated carriers depends on the generation rate (or $J_{\text{gen}}$) with the magnitude of $J/J_{\text{gen}}$ generally decreasing with increasing light intensity. Indeed, Eq. (41) reproduces both the simulated intensity dependence of $J_{\text{SC}}/J_{\text{gen}}$ and *J-V* curve at 1 sun as shown in Fig. 4(b) and (d), respectively. The degree of second-order recombination is critically determined by $K$ as per Eq. (40); at short-circuit, Eq. (40) can be approximated by

$$K \approx \frac{\beta J_{\text{gen}} d^3}{4\mu_n \mu_p V_{bi}^2} \times \left(1 - \frac{\theta kT}{qV_{bi}}\right)^2. \tag{42}$$

Apart from the photogeneration current ($J_{\text{gen}}$), the degree of the second-order recombination also depends on the recombination coefficient and the product of the electron and hole mobility. When $K \ll 1$, corresponding low intensities, small $\beta$, or high enough mobilities, Eq. (41) reduces to Eq. (33) as the second-order recombination within the collection zone is negligible. Conversely, in the high intensity regime, when $K \gg 1$, second-order recombination dominates over charge extraction of the photogenerated charge carriers. In this regime, $J$ approaches a linear $V$ dependence while displaying a sublinear intensity dependence of the form $J \propto G^{1/2}$ (or $J/J_{\text{gen}} \propto G^{-1/2}$), consistent with previous reports [24,44,45].

### D. Validation of the analytical model

To substantiate the presented analytical framework for the current density, we next compare the obtained diode equation (Eq. (41)) against numerical simulations of different cases. Figure 6 shows the numerically simulated *J-V* characteristics for varying mobilities, $\beta$ and mobility imbalance $\mu_p/\mu_n$ under 1 sun and 0.001 suns illumination conditions. Upon comparing the numerically simulated *J-V* curves (coloured solid lines) with the predictions based on Eq. (41) (black dashed lines) an excellent agreement is obtained for voltages sufficiently below $V_{bi}$, in case of balanced mobilities. This is particularly true for the low-intensity cases where a virtually perfect agreement is obtained across the board. At higher forward-bias voltages, however, a deviation is obtained for systems with low mobilities or large $V_{\text{OC}}$. This can be attributed to the fact that the approximation for $F_0$ [Eq. (21)] in Eq. (41) is no longer valid at high voltages (and that Eq. (41) diverges for $d_{\text{eff}} \to 0$). Note that Eq. (41) is essentially indistinguishable from Eq. (33) at 0.001 suns, but also at 1 sun for mobilities $> 10^{-3}$ cm$^2$/Vs in Fig. 4.

A good agreement is also obtained for imbalanced mobilities, especially at low intensities. At large enough mobility imbalances, however, a deviation is eventually observed between the analytically predicted and the simulated *J-V* curves. This deviation is attributed to the inevitable space charge build-up of the slower photogenerated carriers at high intensities, which screens the electric field inside the active layer, resulting in strongly space-charge-limited photocurrents [46,47,48,49].

The developed analytical framework can be used to gain insights about loss mechanisms in organic solar cells and photodiodes. In addition, the derived approximations can be used to extract material parameters from experimental data, for example using a global fitting procedure, in systems where bimolecular recombination is the dominant loss mechanism of photogenerated charge carriers. In this regard, to further validate the analytical model, we applied our diode equation to experimental data of the OPV model system PM6:ITIC [45,50].

The experimental *J-V* curves of a 110 nm thick PM6:ITIC device under 1 sun illumination is depicted in Fig. 7(a). The corresponding normalized $J_{\text{SC}}/J_{\text{gen}}$ is shown as a function of light intensity in Fig. 7(b), assuming $J_{\text{gen}}$ to be proportional to the light intensity. The details of the experiments and device fabrication can be found in elsewhere [45,50]. The carrier mobilities for this system were previously found to balanced and estimated to be $\mu = 2 \times 10^{-4}$ cm$^2$/Vs [45]. As evident from the pronounced intensity dependence of $J_{\text{SC}}/J_{\text{gen}}$, the PM6:ITIC device suffers from second-order bimolecular recombination at the higher intensities (above 1 sun). The solid lines in Fig. 7 indicate the qualitative fit based on Eq. (41) using $\beta$ and $V_{bi}$ as fitting parameters. A good qualitative fit is obtained with $\beta \approx 4 \times 10^{-11}$ cm³/s and $V_{bi} \approx 1.2$ V. The obtained value for $\beta$ is about 7 times smaller than the Langevin recombination coefficient, consistent with the moderate efficiency of this system [50]. The deviation between the experimental data and the fit at the highest intensities in Fig. 7(b) is attributed to additional losses induced by the series resistance of the external circuit (cf. Section III.E).

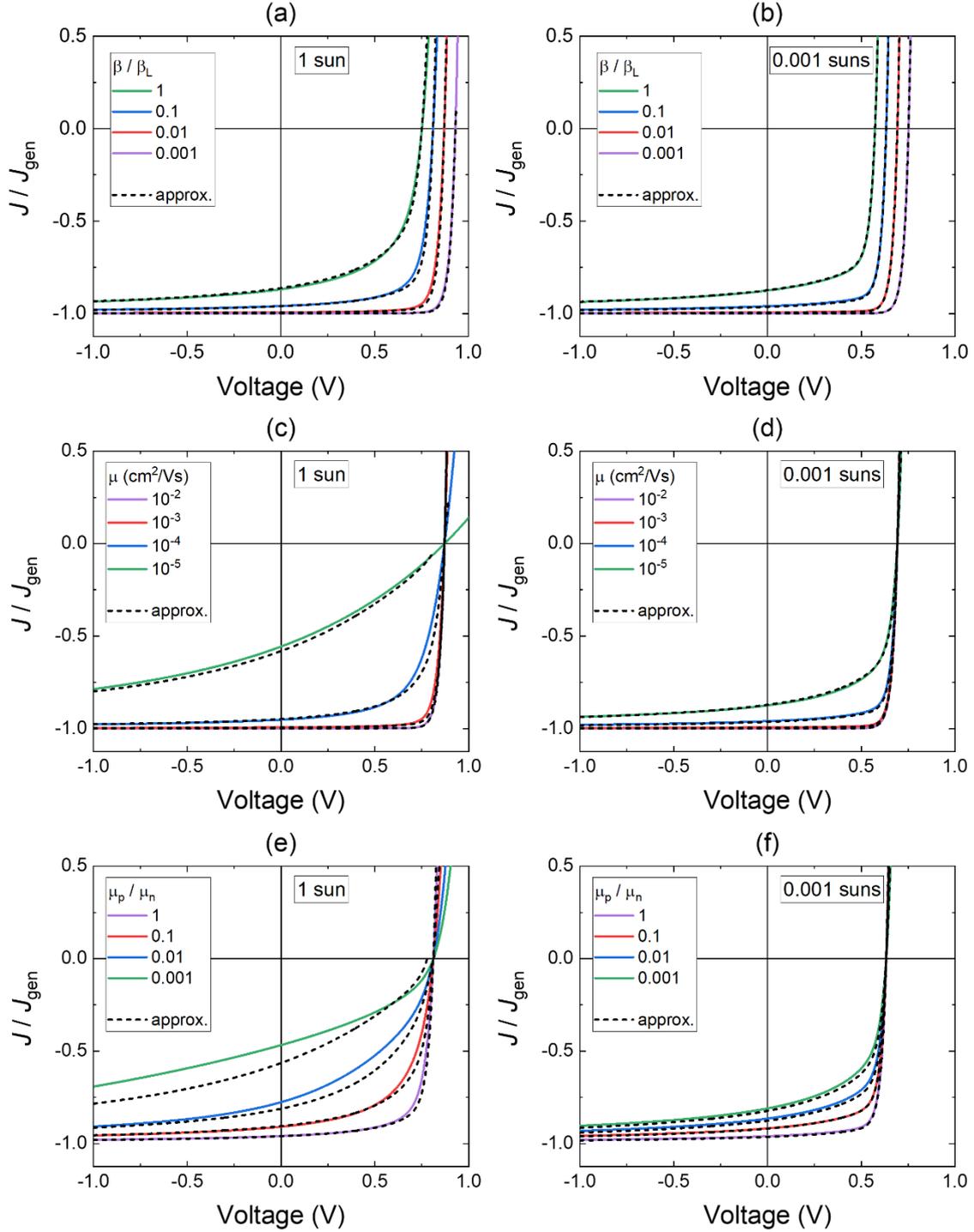

**Figure 6.** Comparison between numerical simulations indicated by the coloured solid lines and the analytical approximations Eq. (41) indicated by the corresponding dashed lines are shown under 1 sun (left-hand-side panel) and 0.001 suns (right-hand-side panel) illumination conditions. In (a) and (b), the simulated *J-V* characteristics for the case of varying $\beta/\beta_L$ are shown for $\mu_n = \mu_p = 10^{-3}$ cm²/Vs. In (c) and (d), the simulated *J-V* characteristics at different but balanced mobilities are shown for $\beta = 10^{-11}$ cm³/s. In (e) and (f), the simulated *J-V* characteristics at increasing mobility imbalance $\mu_p/\mu_n$ is shown for a fixed $\mu_n = 10^{-3}$ cm²/Vs and constant $\beta = 10^{-10}$ cm³/s.

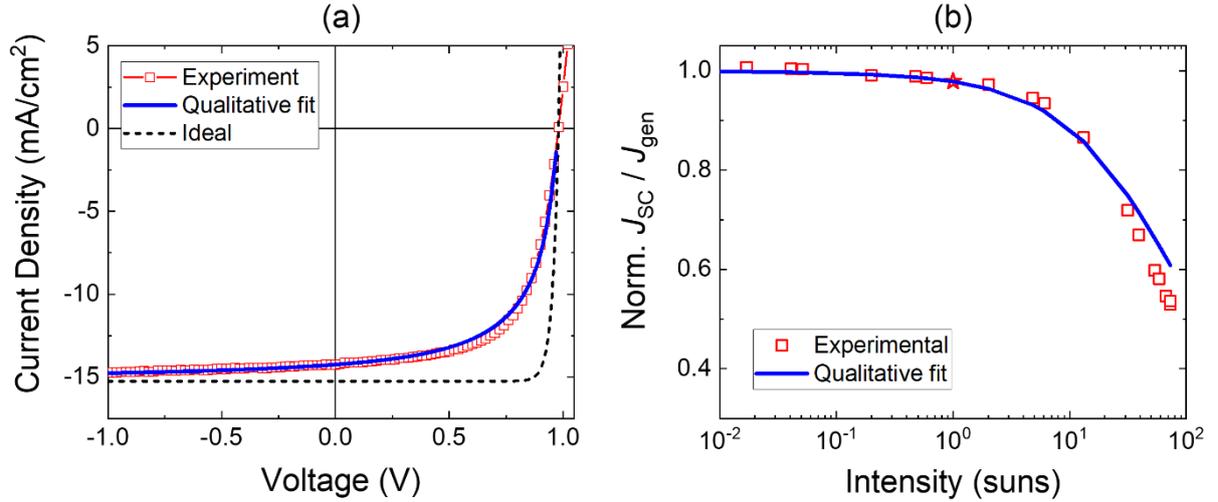

**Figure 7.** (a) Experimental *J-V* curves (symbols) of organic solar cell device based on PM6:ITIC. In (b) normalized $J_{SC}/J_{gen}$ as a function of light intensity for PM6:ITIC (symbols) measured under short-circuit conditions are shown; star-shaped symbol indicate the value at 1 sun. The solid lines in (a) and (b) indicate the corresponding qualitative fits using Eq. (41).

### E. Implications for photovoltaic device performance

The presented theoretical framework implies that the charge collection loss induced by bimolecular recombination can generally be partitioned into zeroth, first and second order contributions. The zeroth order case [Eq. (17)] represents perfect charge collection of photogenerated carriers, containing only fundamental losses associated with the $V_{OC}$. Conversely, first and second-order losses directly reduce the number of collected photogenerated carriers. The corresponding contributions from first- and second-order losses are highlighted by shaded areas in Fig. 4(c) and (d) for the case $\mu = 10^{-4}$ cm²/Vs. These results suggests that even at 1 sun the predominant charge collection loss due to bimolecular recombination in OPVs (with ohmic contacts) is first-order, while second-order losses only show up at low mobilities which may not be relevant to state-of-the art OPVs. The implications of first-order charge collection losses for photovoltaic devices are discussed below.

#### 1. Fill Factor and PCE

The presence of first-order current losses, which depend on the voltage in accordance with Eq. (33), has important implications for the FF, and ultimately the maximal achievable PCE in OPVs. For a solar cell under a given light intensity ($I_L$), the PCE and FF are defined through $\text{PCE} = |J(V_{\text{mp}})V_{\text{mp}}|/I_L =$

$\text{FF} \times J_{\text{SC}} V_{\text{OC}} / I_L$, where $V_{\text{mp}}$ is the voltage at which the output power of the device is maximal. In the ideal diode limit, when $J(V)$ is given by Eq. (17), the FF can be approximated as [28]

$$\text{FF}_{\text{ideal}} = \frac{\frac{qV_{\text{OC}}}{kT} - \ln\left(1 + \frac{qV_{\text{OC}}}{kT}\right)}{1 + \frac{qV_{\text{OC}}}{kT}}, \qquad (43)$$

with the corresponding maximum power point voltage given by

$$V_{\text{mp}}^{\text{ideal}} = V_{\text{OC}} - \frac{kT}{q}\ln\left(1 + \frac{qV_{\text{OC}}}{kT}\right). \qquad (44)$$

Eq. (43) strictly applies for the ideal-diode regime, corresponding to the limit of large mobilities. However, an upper limit for the FF in case of finite mobilities can be estimated assuming that the charge collection is only limited by first-order recombination with injected carriers. Then, noting that the principal voltage dependence in Eq. (33) is determined by the $J_{\text{ideal}}(V)$ component, while the prefactor varies relatively slowly with voltage, we may take $V_{\text{mp}} \approx V_{\text{mp}}^{\text{ideal}}$ as an upper limit estimate for $V_{\text{mp}}$. Subsequently, the FF is reduced relative to the ideal limit by a factor corresponding to the ratio between the prefactor at $V = V_{\text{mp}}$ and at short-circuit as:

$$\text{FF} \approx \frac{\text{FF}_{\text{ideal}}}{\left(1 - \frac{\theta kT}{qV_{bi}}\right)} \times \left(1 - \frac{\theta kT}{qF_0(V_{\text{mp}}^{\text{ideal}})d}\right), \qquad (45)$$

where $\text{FF}_{\text{ideal}}$ and $V_{\text{mp}}^{\text{ideal}}$ is given by Eq. (43) and Eq. (44), respectively.

In accordance with Eq. (45), the FF is expected to depend on the ratios between the bimolecular recombination coefficient and the mobilities, i.e., $\beta/\beta_n$ and $\beta/\beta_p$. Fig. 8 shows the FF as a function of the sum of the ratios, $\beta/\beta_R$, where $\beta_R = \left(\beta_p^{-1} + \beta_n^{-1}\right)^{-1}$. The symbols in Fig. 8(a) and (b) indicate numerically simulated FFs from a wide set of different mobilities, bimolecular recombination coefficients, mobility imbalances, and generation rates, while keeping the $V_{\text{OC}}$ fixed at 0.81 V and 0.63 V, respectively. Note that $\beta$ in these simulations have been restricted to $\beta/\beta_L \leq 1$. The approximation based on Eq. (45) for the special case $\beta_n \gg \beta_p$ (solid lines) and $\beta_n = \beta_p$ (dashed lines), corresponding to the expected upper and lower bounds for Eq. (45) have been included for comparison. Eq. (45) provides a good estimate of the simulated FFs at small $\beta/\beta_R$ (large mobilities and/or small $\beta$) and low $V_{\text{OC}}$, in particular. At larger $\beta/\beta_R$ and higher $V_{\text{OC}}$, on the other hand, non-uniform electric fields and space charge effects in the active layer become increasingly important. Eventually, as $\beta/\beta_R \gg 1$, corresponding to highly imbalanced mobilities, the FF at high $V_{\text{OC}}$ approaches a value close to 0.385 as the simulated devices become severely space-charge-limited.

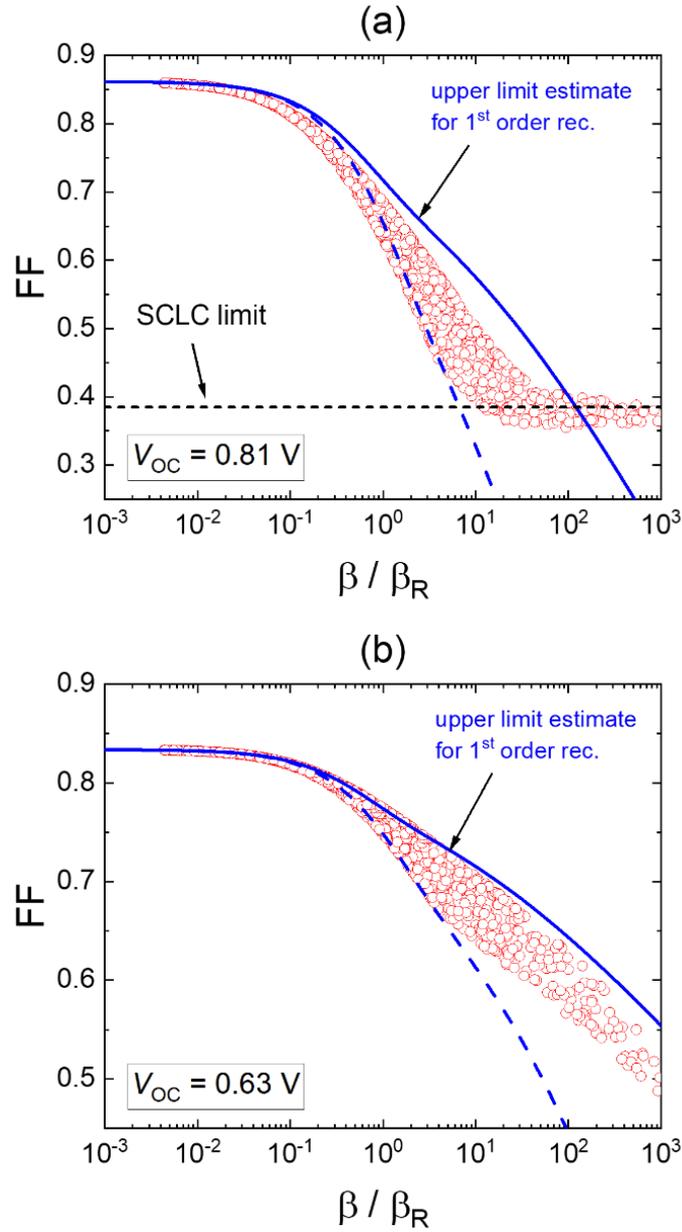

**Figure 8.** The Fill Factor as a function of $\beta/\beta_R$ as extracted from a large set of numerically simulated devices (open symbols) with different $\mu_n, \mu_p, \beta/\beta_L$ and $G$ at fixed open-circuit voltages of (a) $V_{OC} = 0.81$ V and (b) $V_{OC} = 0.63$ V. In the simulations, $\beta/\beta_L$ spans between 0.001 and 1, while $\mu_n$ and $\mu_p$ are allowed to independently vary between $10^{-6}$ and $10^{-2}$ cm$^2$/Vs. For comparison, the approximations based on Eq. (45) for $\beta_n \gg \beta_p$ (solid lines) and $\beta_n = \beta_p$ (blue dashed lines), where the former case corresponds to the estimated upper limit of the FF when the charge collection is limited by first-order recombination with injected carriers. The black dotted line in (a) corresponds to FF = 0.385 expected for space-charge limited photocurrents (SCLC).

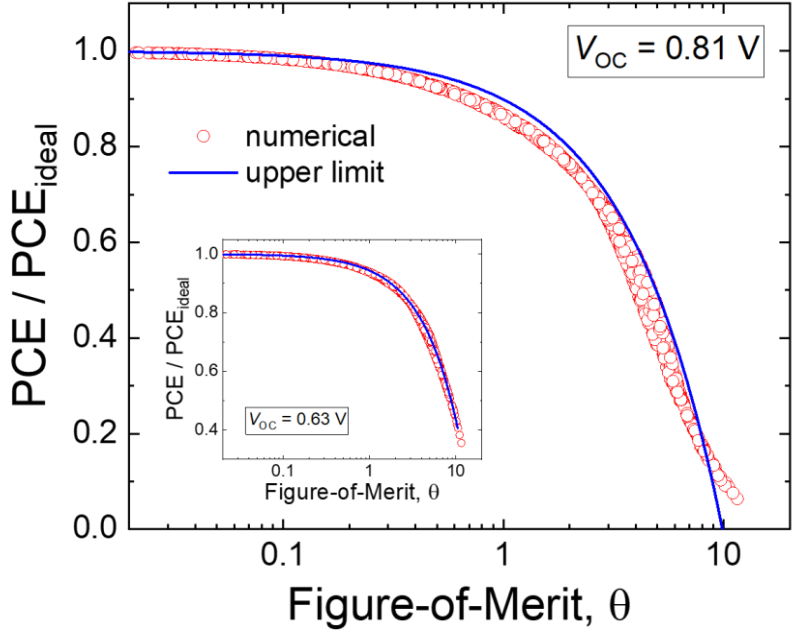

**Figure 9.** The $\text{PCE}/\text{PCE}_{\text{ideal}}$ of the numerically simulated devices from Fig. 8 are shown as a function of the figure-of-merit $\theta$ at $V_{\text{OC}} = 0.81$ V, as indicated by open symbols. The corresponding case for $V_{\text{OC}} = 0.63$ V is shown in the inset. The upper limit estimate of the PCE based on Eq. (46), expected in the presence of first-order recombination with injected carriers, is depicted by the solid lines.

Finally, combining Eq. (45) with the corresponding $J_{\text{SC}}$ predicted by Eq. (33), an upper limit estimate of the PCE for low-mobility thin-film devices with ohmic contacts can be established. The corresponding PCE takes the form

$$\text{PCE} = \text{PCE}_{\text{ideal}} \times \left(1 - \frac{\theta kT}{qF_0(V_{\text{mp}}^{\text{ideal}})d}\right), \tag{46}$$

where $\text{PCE}_{\text{ideal}}$ is the PCE expected in case of perfect charge collection. The parameter $\theta$ may thus be taken as an associated figure-of-merit for charge collection. Fig. 9 shows the $\text{PCE}/\text{PCE}_{\text{ideal}}$ of the numerically simulated devices from Fig. 8 but plotted as a function of $\theta$, where $\theta$ is obtained from Eq. (34) using $\beta/\beta_n$ and $\beta/\beta_p$ as input. As shown, for a given $V_{\text{OC}}$ (and $V_{bi}$) the simulated data set coalesces into a common trend which only depends on $\theta$. Indeed, comparing the simulated trend with the upper limit estimate Eq. (46), a very good overall agreement can be obtained. As a result, Eq. (46) can be used to separately estimate the loss in the PCE associated with charge collection in OPV devices with selective ohmic contacts. These results can also be used to facilitate machine-learning for PCE predictions. Finally, we note that $\theta$ [Eq. (34)] is markedly different to previously proposed figures-of-merit which are proportional to $\frac{\beta G}{\mu_n \mu_p}$ [13,14,22]. The fact that the spread in the numerically simulated data (from Fig. 8) is drastically reduced when parametrized against $\theta$ (see Fig. 9) elevates $\theta$ as a more accurate figure-of-merit for charge collection in thin-film OPVs.

## 2. The dark current and the reciprocity between charge injection and extraction

The theoretical framework based on Eq. (33) can be consolidated with the principle of reciprocity between charge injection and extraction [51], provided that the voltage dependence of the photovoltaic external quantum efficiency (EQE) is accounted for [52]. In general, the EQE describes the efficiency for incident photons to be converted into collected electron-hole pairs at the contacts. In excitonic semiconductors, the EQE can be written as $\text{EQE} = \eta_{\text{abs}} \eta_{\text{CGY}} \eta_{\text{col}}$, where $\eta_{\text{abs}}$ is the efficiency for an incident photon to be absorbed and generate an exciton and $\eta_{\text{CGY}}$ is the efficiency for a photogenerated exciton to generate a free electron-hole pair. Finally, $\eta_{\text{col}}$ is the charge collection efficiency of photogenerated carriers defined as $\eta_{\text{col}} = |J - J_{\text{dark}}|/J_{\text{gen}}$, where $J_{\text{dark}}$ is the current density under dark conditions. Note that, from this perspective, $J_{\text{gen}}$ can be equivalently expressed as $J_{\text{gen}} = q \int_0^\infty \eta_{\text{abs}} \eta_{\text{CGY}} \Phi_{\text{ext}} \, dE$, where $\Phi_{\text{ext}}(E)$ is the photon flux of the external light source used to photogenerated charge carriers (e.g., 1 sun).

In accordance with Eq. (33), under conditions when recombination between photogenerated and injected charge carriers dominates, the charge collection efficiency is given by

$$\eta_{\text{col}}(V) = 1 - \frac{\theta k T}{q F_0(V) d}, \tag{47}$$

for $V \ll V_{bi}$. Based on Eq. (47), $\eta_{\text{col}}$ depends on the applied voltage but is independent of light intensity, consistent with the first-order nature of the current loss. In fact, Eq. (47) applies to dark conditions as well, noting that the current density Eq. (33) in the dark is given by

$$J_{\text{dark}}(V) = \left(1 - \frac{\theta k T}{q F_0(V) d}\right) \times J_0 \left[\exp\left(\frac{qV}{kT}\right) - 1\right], \tag{48}$$

for $V \ll V_{bi}$. Hence, the effect of finite mobilities, is to reduce the injected dark current density, relative to the ideal-diode case $J_{\text{ideal}} = J_0 \left[\exp\left(\frac{qV}{kT}\right) - 1\right]$ (i.e., Eq. (17) in the dark), by the prefactor $\eta_{\text{col}}(V)$ given by Eq. (47). Concomitantly, $J_{\text{dark}}(V)$ can be described in terms of an effective dark saturation current density $J_{0,\text{eff}}(V) = \eta_{\text{col}}(V) \times J_0$, which depends on the applied voltage. This is demonstrated in Fig. 10, where *J-V* curves under dark conditions are simulated and compared to Eq. (48), showing excellent agreement for $V \ll V_{bi}$.

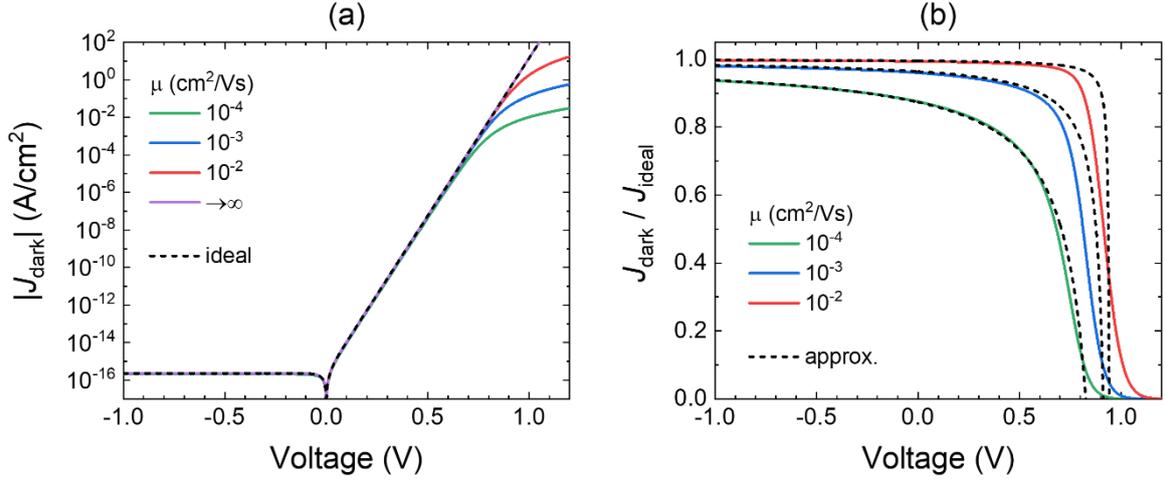

**Figure 10.** (a) Simulated dark current densities at different mobilities with $\beta$ of $10^{-10}$ cm$^3$/s. (b) Normalized dark current densities relative to the idealized expected dark current density [Eq. (17)] in the limit of high mobilities. The approximations based on Eq. (48) are indicated by dashed lines.

Under illumination, the rates of charge carrier injection and extraction are exactly balanced at open circuit (where $J = 0$). From Eq. (33), the open-circuit voltage is given by

$$V_{\text{OC}} = \frac{kT}{q} \ln\left(1 + \frac{J_{\text{gen}}}{J_0}\right). \tag{49}$$

However, since $J_0$ is defined as the (saturated) thermal generation current density in the bulk, and noting that charge carriers can be thermally generated either radiatively or non-radiatively, we can write $J_0 = (q/\eta_{\text{rad}}) \int_0^\infty \eta_{\text{abs}} \eta_{\text{CGY}} \Phi_{\text{BB}}\, dE$, where $\Phi_{\text{BB}}(E)$ is the Black-body photon flux of the environment and $\eta_{\text{rad}}$ is the radiation efficiency. On the other hand, provided that $\eta_{\text{col}}$ remain independent of intensity, it follows that $J_{\text{gen}}/J_0 = \eta_{\text{col}}(V) J_{\text{gen}}/J_{0,\text{eff}}(V)$. If the EQE is evaluated at low intensities under short-circuit conditions ($V = 0$), Eq. (49) can then be reformulated as

$$V_{\text{OC}} = \frac{kT}{q} \ln\left(1 + \frac{J_{\text{SC}}^*}{J_{0,\text{SC}}}\right), \tag{50}$$

consistent with detailed balance, where $J_{\text{SC}}^* = q \int_0^\infty \text{EQE}(E) \phi_{\text{ext}}(E)\, dE$ is the expected $J_{\text{SC}}$ in the absence of second-order losses, while $J_{0,\text{SC}} = (q/\eta_{\text{rad}}) \int_0^\infty \text{EQE}(E) \phi_{\text{BB}}(E)\, dE$ corresponds to $J_{0,\text{eff}}(V = 0)$. Note that $J_{0,\text{SC}}$ generally differs from $J_0$. It should be emphasized that Eq. (50) assumes that $J_{\text{SC}}^* = J_{\text{SC}}$ which only apply for conditions when Eq. (42) is negligibly small, i.e., $K \ll 1$. For very low mobilities, this condition is no longer valid, and a deviation from reciprocity is expected at open circuit, in agreement with previous findings [52].

**E. Other effects**

The presented diode equations neglect the effect of external resistive losses. In real devices, the voltage $V$ over the active layer (diode) is generally different to the externally applied voltage $V_{\text{ext}}$, due to the finite resistance of the wires (connecting the load to the diode) and the electrodes. In addition, diode devices suffer from parasitic shunts, commonly attributed to extrinsic non-idealities from the device fabrication. The device-intrinsic current density $J(V)$ (considered above) is related to the external current density $J_{\text{meas}}$, measured as a function of $V_{\text{ext}}$, via

$$J_{\text{meas}}(V_{\text{ext}}) = J(V) + \frac{V}{\mathcal{R}_{\text{shunt}}}, \tag{51}$$

where $V = V_{\text{ext}} - J_{\text{meas}}\mathcal{R}_{\text{series}}$. Here, $\mathcal{R}_{\text{series}}$ is the combined series resistance (in units of Ωcm²) of the external circuit and the electrodes and $\mathcal{R}_{\text{shunt}}$ is the shunt resistance (in Ωcm²). While voltage losses induced by a finite series resistance become prevalent when the magnitude of the current density is large (high intensity and/or high forward bias), the effect of the shunt is generally important at low current densities and light intensities (dominating the measured dark current density in the reverse bias).

The presented theoretical treatment for $J(V)$ applies to undoped thin-film diode devices with ohmic contacts, assuming bimolecular recombination to be the dominant recombination mechanism. This corresponds to situations where additional loss channels induced by trap-assisted recombination and surface recombination of minority carriers at the contacts are negligible. If there is a sufficiently large number of traps in the active layer, however, the resultant trap-assisted recombination generally gives rise to an increased charge collection loss [45,53,54,55], which depends on the trap energy and distribution. Similar to doping in the active layer [56], trapped charge carriers may also induce considerable electric field screening inside the active layer [53,57], further reducing the photocurrent. Additionally, if the contacts are non-selective for the extraction of majority carriers, additional charge collection losses and reductions in the open-circuit voltages caused by surface recombination of minority carriers may occur. This loss is expected to be particularly prominent in case of energetically unoptimized contacts and/or large carrier mobilities [58,59,60,61].

Finally, it should be stressed that the theoretical analysis is limited to thin active layers with relatively uniform charge carrier generation rates. While the absorption rate of photons inside the active layer is decays exponentially with the distance from the transparent electrode, the combined effect of optical interference and back reflection from the reflective counter electrode results in smooth generation rates in (optically) thin active layers. For thick active layers, however, the photon absorption rate in the active layer is concentrated near the transparent contact resulting in a highly non-uniform charge carrier generation and asymmetric charge extraction [62,63,64,65]. Furthermore, thick active layers are generally more sensitive to space charge effects induced by photogenerated carriers, trapped charge carriers, or (unintentional) dopants.

## 4. CONCLUSIONS

In conclusion, we have presented an analytic framework for describing the device physics of charge collection in thin-film OPVs with ohmic contacts. Based on this framework, a diode equation describing the *J-V* characteristics is derived, showing excellent agreement with numerical simulations. The presented framework is further employed to analytically understand what limits the charge collection efficiency, FF, and ultimately the PCE in organic solar cells. The obtained findings provide vital physical insights into the interplay between charge carrier extraction, injection, and bimolecular recombination and how these processes influence the device performance of OPV devices. The developed diode equations are not limited to organics but apply generally to sandwich-type thin-film devices based on low-mobility semiconductors with ohmic contacts.

[42] For $x > x^*$, Eq. (4) is approximated as $\frac{\partial^2 \phi(x)}{\partial x^2} \approx \frac{qn_0(x)}{\varepsilon\varepsilon_0}$, where $n_0(x)$ is given by Eq. (14); in this region, we then find $\phi(x) = V_{bi,0} - V - \frac{2kT}{q}\ln\left(\frac{2kT}{qF_0\lambda_{\text{cat}}}\sinh\left[\frac{qF_0(d-x)}{2kT} + \sinh^{-1}\left(\frac{qF_0\lambda_{\text{cat}}}{2kT}\right)\right]\right)$, where $\lambda_{\text{cat}} = \sqrt{2\varepsilon\varepsilon_0 kT/[q^2 n_{\text{cat}}]}$. Equating this expression with Eq. (19) at $x = x^*$ in case of ohmic contacts ($\lambda_{\text{an}}, \lambda_{\text{cat}} \ll d$) yields $V_{bi,\text{eff}} = V_{bi,0} - \frac{2kT}{q}\ln\left(\frac{q^2 d^2 \sqrt{p_{\text{an}} n_{\text{cat}}}}{2\varepsilon\varepsilon_0 kT}\right) + \frac{4kT}{q}\ln\left(\frac{q[V_{bi,\text{eff}}-V]}{kT}\right)$ for small voltages, where $V_{bi,\text{eff}} \equiv F_0 d + V$. Approximating $V_{bi,\text{eff}}$ by applying Newton's iteration method once with $V_{bi}$ as the initial guess for $V_{bi,\text{eff}}$, we find $V_{bi,\text{eff}} \approx V_{bi} + \frac{4kT}{q}\ln\left(1 - \frac{V}{V_{bi}}\right)\left(1 - \frac{4kT}{q[V_{bi}-V]}\right)^{-1}$. Finally, we approximate $F_0$ by making use of a linear approximation of $V_{bi,\text{eff}}$ between $V = 0$ and $V = V_{bi}/2$ such that $F_0 d \approx V_{bi} - \left[V_{bi,\text{eff}}(0) - V_{bi,\text{eff}}\left(\frac{V_{bi}}{2}\right) + V_{bi}/2\right] \times (2V/V_{bi})$.

[43] A correction from diffusion is estimated as $\Delta J_{n,\text{diff}} = \mu_n kT \left[\frac{\partial \Delta n}{\partial x}\big|_{x=d_{1/2}} - \frac{\partial \Delta n}{\partial x}\big|_{x=x_{Rn}}\right]$, where $d_{1/2} = (d + x_{Rn} - x_{Rp})/2$ is the midplane of the collection zone. Using Eq. (36), one finds $\Delta J_{n,\text{diff}} =$

$J_{\text{gen}}^* \times \left(\frac{J^2 kT}{J_\beta^2 q F_0 d_{\text{eff}}}\right)$. The corrected current density thus reads $J_{\text{corr}} = J + \Delta J_{n,\text{diff}}$, where $J$ corresponds to the uncorrected current density. On the other hand, expanding the left-hand-side of Eq. (37), and solving for $J$ yields $J = J_{\text{gen}}^*/\sqrt{1 + (J_{\text{gen}}^*/J_\beta)^2}$. Hence, after combining everything together we find

$J_{\text{corr}} = \frac{J_{\text{gen}}^*}{\sqrt{1+(J_{\text{gen}}^*/J_\beta)^2}} \times \left[1 - \frac{kT}{qF_0 d_{\text{eff}}} \frac{(J_{\text{gen}}^*/J_\beta)^2}{\sqrt{1+(J_{\text{gen}}^*/J_\beta)^2}}\right]$. Finally, assuming the correction term to be much smaller than unity, we can approximate $J_{\text{corr}} \approx J_{\text{gen}}^*/\sqrt{1 + (J_{\text{gen}}^*/J_{\beta,\text{corr}})^2}$, where $J_{\beta,\text{corr}}$ is given by Eq. (38).

**ACKNOWLEDGEMENTS**

This work was funded by the UKRI through the EPSRC Program Grant EP/T028513/1 Application Targeted and Integrated Photovoltaics.